\documentclass{ws-procs10x7}
\usepackage{balance}

\newcolumntype{d}[1]{D{.}{.}{#1}}

\makeindex
\begin{document}

\title{SEARCH FOR SINGLE-TOP-QUARK PRODUCTION AT THE TEVATRON}

\author{WOLFGANG WAGNER$^*$}

\address{Institut f\"ur Experimentelle Kernphysik, Universit\"at Karlsruhe, 76128 Karlsruhe, Germany\\$^*$E-mail: wagner@ekp.uni-karlsruhe.de}

\address{for the CDF and D\O \ Collaborations}


\twocolumn[\maketitle\abstract{This article reports on recent 
searches for single-top-quark production by the CDF and 
D\O \ collaborations at the Tevatron. 
Neither two CDF analyses, using data corresponding to
$695\,\mathrm{pb^{-1}}$ of integrated luminosity, nor a D\O \ 
search based on
$370\,\mathrm{pb^{-1}}$ of data can establish a signal for this standard model
process. These null results translate in upper limits on the
cross section of $2.9\,\mathrm{pb}$ for the $t$-channel production
mode and $3.2\,\mathrm{pb}$ for the $s$-channel mode
at the 95\% C.L.
In addition,
the D\O \ collaboration has searched for non-standard model
production of single-top-quarks via a heavy $W^\prime$ boson. 
No signal of this process is found resulting in lower mass limits
of $610\,\mathrm{GeV}/c^2$ for a left-handed $W^\prime$ and
$630\;\mathrm{GeV}/c^2$ for a right-handed $W^\prime$ boson.  
}
\keywords{single-top; electroweak top quark production; $W^\prime$ boson.}
]

\def\EtMiss  {E_\mathrm{T}\!\!\!\!\!\!\!/ \ \; }
\def\EtMissVec {{E}_\mathrm{T}\!\!\!\!\!\!\!/ \;\;}

\section{Introduction}
According to the standard model, in $p\bar{p}$ collisions at the 
Tevatron top 
quarks can be created in pairs via the strong force, of singly via the
electroweak interaction. The latter production mode is referred to as 
``single-top-quark'' production and takes place mainly through the 
$s$- or 
$t$- channel exchange of a $W$ boson. 
The CDF and D\O~
collaborations have published single-top results at 
$\sqrt{\mathrm{s}}=1.8$ 
TeV and $\sqrt{\mathrm{s}}=1.96$ TeV~\cite{allcdf,alldzero}. None
of these analyses established single-top evidence, and 95\%
confidence level (C.L.) upper limits on the single-top production cross
section were set.

The theoretical single-top production cross section is 
$\sigma_{s+t}=2.9\pm 0.4$ pb for a top mass of 175 GeV/$c^{2}$~\cite{theo}.
Despite this small rate, the main obstacle in finding
single-top is in fact the large associated background. After
all section requirements are imposed, the signal to
background ratio is approximately 1/20. This challenging,
background-dominated dataset is the main motivation for using
multivariate techniques.

\section{Standard Model Searches}
In this article we present three new searches, two by CDF~\cite{cdfnew}, in
subsections \ref{sec:nn} and \ref{sec:cdfll}, and one by 
D\O~\cite{d0new}, in subsection \ref{sec:d0lr}. 
The two CDF analyses use a dataset corresponding to
$695\,\mathrm{pb^{-1}}$. One analyses is performed with neural networks,
the second utilizes likelihood functions. Both use a common event
selection.

\subsection{CDF Neural Network Search}
\label{sec:nn}
The CDF event selection exploits the kinematic features of the signal
final state, which contains a top quark, a bottom quark, 
and possibly additional light quark jets. 
To reduce multijet backgrounds, the $W$ originating from the top
quark is required to have decayed leptonically. 
One therefore demands a single high-energy
electron or muon ($E_{T}(e)>20$ GeV, or $P_{T}(\mu)>20$ GeV/$c$) 
and large missing transverse energy from the undetected 
neutrino $\EtMiss$\/$>$20 GeV. 
The analysis uses electrons measured in the central and in the 
forward calorimeter. The usage of forward electrons is new for
top physics analyses at CDF.
The remaining backgrounds belong to the following categories: 
$Wb\bar{b}$, $Wc\bar{c}$, $Wc$, mistags (light quarks 
misidentified as heavy flavor jets), non-$W$ (events where a jet is 
erroneously identified as a lepton), and diboson $WW$, $WZ$, and $ZZ$.
We remove a large fraction of the backgrounds by demanding exactly 
two jets with $E_{T}>15$ GeV and $|\eta|<2.8$ be present in the event. 
At least one of these two jets should be tagged as a $b$ quark jet 
by using displaced vertex information from the silicon vertex 
detector (SVX).
The non-$W$ content of the selected dataset is further reduced 
by requiring the
angle between the $\EtMissVec$ vector and the 
transverse momentum vector of the leading
jet to satisfy: $0.5<\Delta \Phi< 2.5$.
The numbers of expected and observed events are listed 
in table~\ref{tab:nnexpect}. 
\begin{table}
\tbl{Expected number of signal and background events 
and total number of events observed in 695 pb$^{-1}$ in the
CDF single-top dataset.\label{tab:nnexpect}}
{\begin{tabular}{lc}
\hline
Process                 & $N$ events \\
\hline
$t$-channel              & $16.7\pm1.7$ \\
$s$-channel              & $11.5\pm0.9$ \\
$t\bar{t}$               & $40.3\pm3.5$ \\
diboson, $Z$             & $17.2\pm0.8$ \\
$W+b\bar{b}$             & $170.7\pm 49.2$ \\                
$W+c\bar{c}$             & $64.5\pm 17.3$ \\
$Wc$                     & $69.4\pm 15.3$ \\
$W+q\bar{q}$, mistags    & $164.3\pm 29.6$ \\
non-$W$                  & $119.5\pm 40.4$ \\
\hline
Total                    & $674.1\pm 96.1$ \\
\hline
Observed                 & 689 \\
\hline
\end{tabular}}
\end{table}

\begin{figure}[!th]  
\begin{center}
\includegraphics[width=0.40\textwidth]{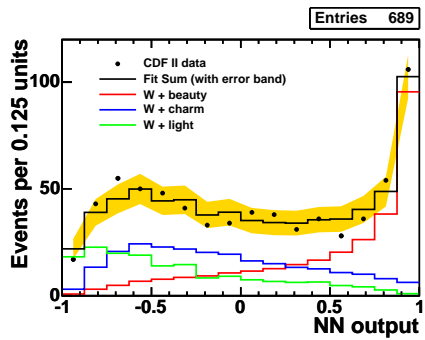}
\end{center}
\caption[nnbtag]{\label{fig:nnbtag}Output distribution of the neural
net $b$ tagger for 689 candidate events in the $W+2$ jets bin.
Overlayed are the fitted components of beauty-like, charm-like
and mistag templates.} 
\end{figure}

Using a neural network 14 kinematic or event shape variables 
are combined to a 
powerful discriminant. 
One of the variables is the output of a neural net $b$ tagger.
In figure~\ref{fig:nnbtag} the distribution of this $b$ tag
variable is shown for the 689 data events. In case of double-tagged
events the leading $b$ jet (highest in $E_T$) is included in this
distribution.
The neural net $b$ tagger gives an additional handle to reduce the
large background components where no real $b$ quarks are contained,
mistags and charm-backgrounds. Both of them amount to about 50\%
in the $W+2$ jets data sample even after imposing the requirement that one jet
is identified by the secondary vertex tagger of CDF~\cite{secvtx}.

\begin{figure}[!th]  
\begin{center}
a) \hspace*{0.35\textwidth} \\
\includegraphics[width=0.33\textwidth]{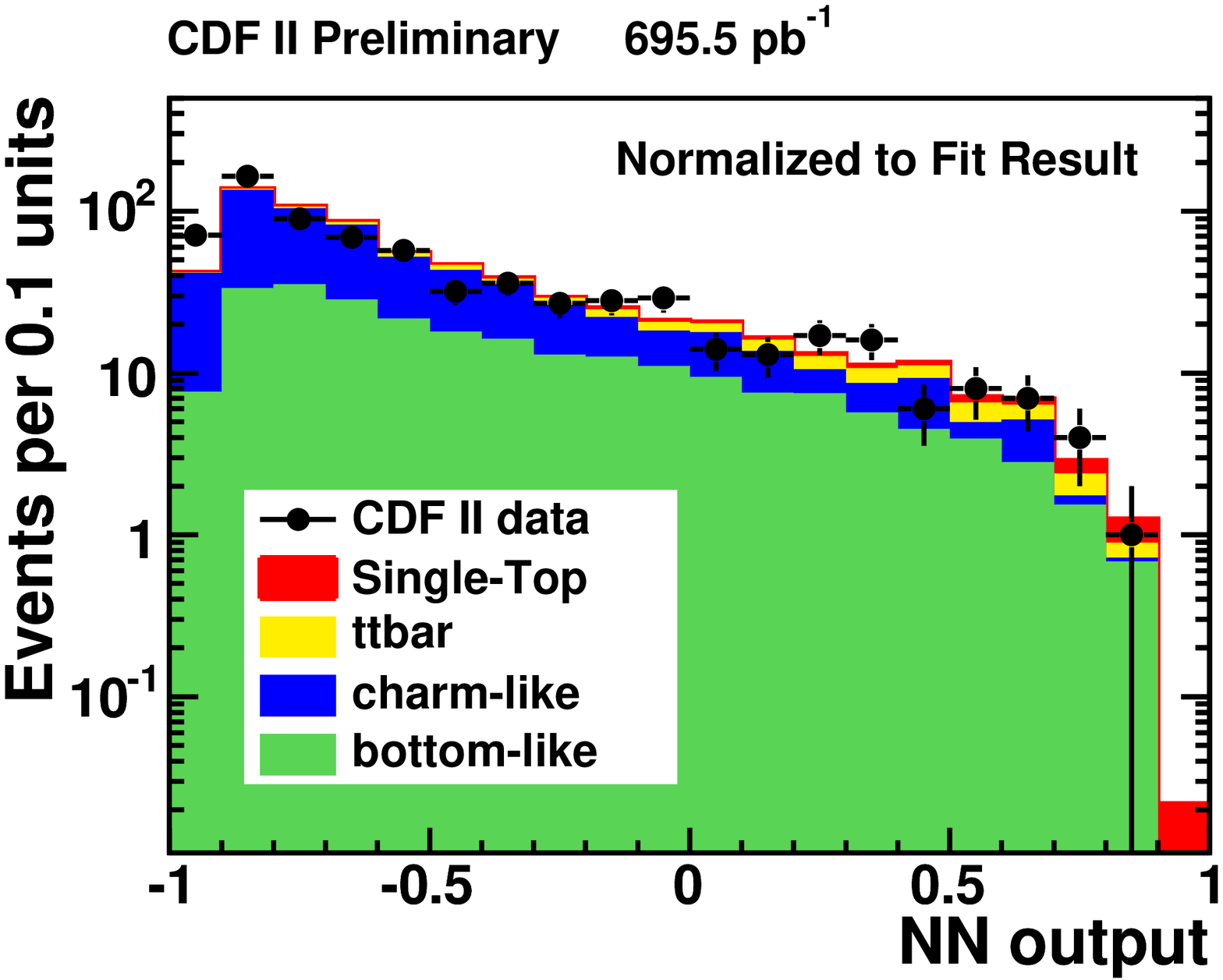} \\
b) \hspace*{0.35\textwidth} \\
\includegraphics[width=0.33\textwidth]{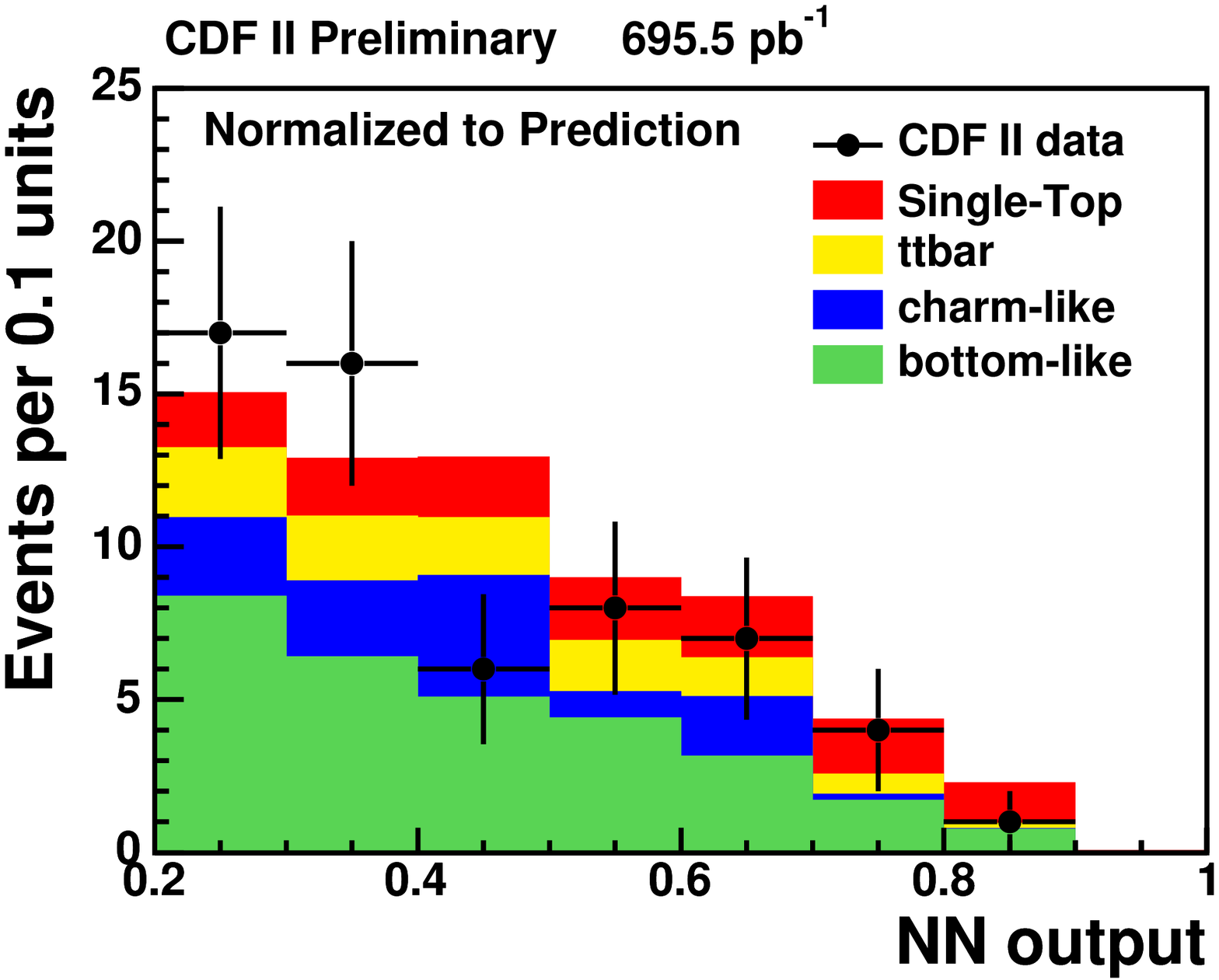}
\end{center}
\caption[nnstdata]{\label{fig:nnstdata}Single-top search with 
  neural networks at CDF:
  a) data compared to the fit result, b) data compared to the 
  standard model expectation in the signal region.} 
\end{figure}

\begin{figure}[!th]
\begin{center}   
\includegraphics[width=0.33\textwidth]{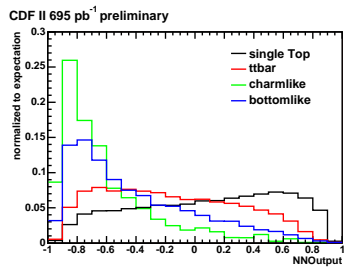}
\end{center}
\caption[nnsttemp]{\label{fig:nntemplates}Monte Carlo template 
 distributions for the single-top neural network search. 
 The distributions are normalized to unit area.}
\end{figure}

Figure~\ref{fig:nnstdata} shows the 
observed data compared to the fit result (a) and the expectation 
in the signal region (b) for the single-top 
neural network.
For comparison, the Monte Carlo
template distributions normalized to unit area are 
shown in figure~\ref{fig:nntemplates}.
The data are fitted with a binned likelihood function.
The $t$- and the $s$-channel are treated as one single-top signal
assuming the ratio of the two processes to be the one predicted by
the standard model. The most probable value of the likelihood function
is $0.8^{+1.3}_{-0.8}\,(\mathrm{stat})\,^{+0.2}_{-0.3}\,(\mathrm{syst})\;\mathrm{pb}$. 
providing no significant evidence for single-top production. 
The corresponding upper limit on the cross section is
$3.4\,\mathrm{pb}$ at the 95\% confidence level, quite close to 
the expected standard model value of $2.9\pm0.4\,\mathrm{pb}$.

To separate $t$- and $s$-channel production two additional networks are
trained and a simulanteous fit to both discriminants is performed.
The fit results are summarized in table~\ref{tab:2dnnResults}.
\begin{table}[!th]
\tbl{Fit results on the separate search for $t$- and $s$-channel 
  single-top production. The quoted limits are set at the 95\%
  C.L.\label{tab:2dnnResults}} 
{\begin{tabular}{lcc}
\hline              
                    & $t$-channel & $s$-channel \\ \hline
\begin{minipage}{0.12\textwidth}
  \vspace*{1mm}
  Observed most probable value \\ \vspace*{1mm}
\end{minipage}& 
\begin{minipage}{0.12\textwidth}
$0.6^{+1.9}_{-0.6}\,(\mathrm{stat})$ \\
$\;\ \ \ ^{+0.1}_{-0.1}\,(\mathrm{sys})\;\mathrm{pb}$
\end{minipage}
&
\begin{minipage}{0.12\textwidth}
$0.3^{+2.2}_{-0.3}\,(\mathrm{stat})$ \\
$\;\ \ \ ^{+0.5}_{-0.3}\,(\mathrm{sys})\;\mathrm{pb}$
\end{minipage} \\
\begin{minipage}{0.12\textwidth}
Observed upper limit \\ \vspace*{1mm}
\end{minipage} & 3.1 pb & 3.2 pb \\
\begin{minipage}{0.12\textwidth}
Expected upper limit \\ \vspace*{1mm} \end{minipage} & 4.2 pb & 3.7 pb \\
\hline
\end{tabular}}
\end{table}
The expected limits are calculated from pseudo-experiments which include 
single-top quark events at the standard model rate.
Again, there is no evidence for single-top production yet.

\subsection{CDF Likelihood Function Analysis}
\label{sec:cdfll}
\begin{figure}[tbh]
\begin{center}
\includegraphics[width=0.40\textwidth]{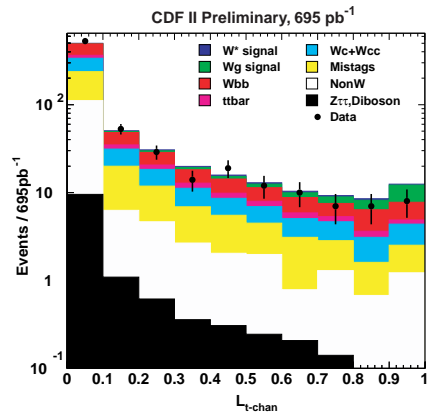}
\caption{\label{fig:tchanlikeData}The distributions of the $t$-channel
  likelihood function for CDF data compared to the Monte Carlo distribution
  normalized to the expected contributions. A logarithmic scale is used.}
\end{center}
\end{figure}

In a second analysis CDF uses likelihood functions to combine several
variables to a discriminant to separate single-top events from background events.
The same data events as for the neural network search are analyzed. 
One likelihood function is defined for the $t$-channel, one for the 
$s$-channel search. Seven or six variables are used, respectively.
The likelihood functions are constructed by first forming histograms of each 
variable. The histograms are produced separately for signal and several background
processes. The histograms are normalized such that the sum of their bin contents
equals 1. For one variable the different processes are combined by computing the
ratio of signal and the sum of the background histograms. These ratios are 
multiplicatively combined to form the likelihood functions. The $t$-channel likelihood
function is shown in figure~\ref{fig:tchanlikeData}.
The observed data show no indication of a single-top signal and are 
compatible with a background-only hypothesis. The $t$-channel analysis
yields an upper limit on the cross section of 2.9 pb at the 95\% 
C.L. A fit to the $s$-channel likelihood function lead to an upper
limit of 5.1~pb. Both likelihood functions are also combined in a 
two-dimensional fit resulting in an upper limit of 4.3 pb on the
combined single-top cross section. The expected upper limit for the
combined search is 3.4 pb, assuming no single-top events to be present.

\subsection{D\O~ Likelihood Ratio Analysis} 
\label{sec:d0lr}
In search for single-top the D\O \ collaboration has analyzed a data set
corresponding to $370\,\mathrm{pb^{-1}}$ of integrated luminosity.
The event selection asks for one isolated electron or muon with 
$p_T > 15\,\mathrm{GeV}$. Electrons are accepted in the pseudorapidity
region of $|\eta|<1.1$, muons in the interval $|\eta|<2.0$. The missing 
transverse energy is required to be above 15 GeV. The D\O \ analysis uses
events with 2, 3 or 4 jets with $E_T>15\,\mathrm{GeV}$ and $|\eta|<3.4$.
The leading jet is required to have a minimum transverse energy of 25 
GeV. At least one jet is required to be identified as originating from
a $b$ quark. A jet lifetime probability algorithm~\cite{d0JLIP} is
used for that purpose. 
After all selection cuts D\O \ observes 443 events, 367 of 
those have a single $b$ tagged jet, 76 events have at least two $b$ tagged
jets. According to the standard model prediction 15.0 $t$-channel and
9.5 $s$-channel events are expected to be present in this data set. 
The expected background is 452 events. 

\begin{figure}[tbh]
\begin{center}
\includegraphics[width=0.40\textwidth]{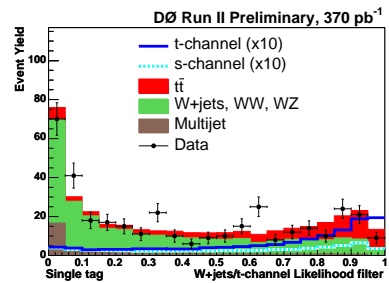}
\caption{\label{fig:tchanWjetsll}The distribution of the $t$-channel
  versus $W$ + jets likelihood discriminant of the D\O \ analysis.}
\end{center}
\end{figure}

The data set is subdivided into the electron and muon channels, and one-tag
and more-than-two-tag samples. Likelihood discriminants are formed to 
separate $t$- and $s$-channel single-top on one hand and 
$t\bar{t}$ and $W$ + jets on the other hand. In total, this approach leads
to 16 likelihood discriminants. 
An example for a likelihood discriminant is shown in 
figure~\ref{fig:tchanWjetsll}.
The combination of all discriminants yields
no evidence for single-top production and is translated into upper
limits on the single-top cross sections at the 95\% C.L.:
$\sigma$ ($t$-channel) $< 4.4$ pb, $\sigma$ ($s$-channel) $< 5.0$ pb.
The corresponding expected limits are 4.3 pb and 3.3 pb, respectively,
assuming no single-top events to be present. 

\section{Search for a $W^\prime$ boson at D\O}
Based on a similar event selection as discussed in section~\ref{sec:d0lr} 
the D\O \ collaboration has also searched for a $W^\prime$ boson in the
decay channel, $W^\prime\rightarrow t\bar{b}$~\cite{d0Wprime}. 
Data corresponding to an integrated luminosity of $230\,\mathrm{pb^{-1}}$
are used.
The signal is modeled
using the event generator {\sc comphep}~\cite{comphep} which takes 
the interference between
a left-handed $W^\prime$ boson and the standard model $W$ boson into 
account.
The analysis is designed to search for a left-handed and a right-handed 
$W^\prime$ boson. The invariant mass of charged lepton, the reconstructed
neutrino and the two leading jets is used as a discriminant.
No evidence for a $W^\prime$ boson is found. Lower limits on the mass 
of the $W^\prime$ are set: $M(W^\prime_L) > 610\,\mathrm{GeV}$ for a 
left-handed $W^\prime$ and $M(W^\prime_R) > 630\,\mathrm{GeV}$ for a
right-handed $W^\prime$. These limits assume that the $W^\prime$ is allowed
to decay to both leptons and quarks.

\section{Conclusions}
The CDF and D\O \ collaborations have performed searches for single-top
production via the electroweak interaction. No evidence for this process
has been found so far. The two CDF analyses are based on a data set
corresponding to an integrated luminosity of $695\,\mathrm{pb^{-1}}$.
Both analyses observe a deficit of events in the signal region of their
discriminant, where single-top events are expected.
Upper limits on the single-top cross sections, quite close to the predicted
values, are set:
$\sigma$($t$-channel)$< 2.9\,\mathrm{pb}$ and
$\sigma$($s$-channel)$< 3.2\,\mathrm{pb}$, both
at the 95\% C.L. Currently data sets corresponding to $1\;\mathrm{fb}^{-1}$ are being
analyzed. Even larger data samples will be available in the future.
It will be very interesting to see whether the trend to a deficit in single-top
events continues or whether first hints of a signal emerge.


\begin{thebibliography}{99}

\bibitem{allcdf} D.~Acosta {\it et al.} (CDF Collaboration), 
{\it Phys. Rev.} {\bf D65}, 091102 (2002); 
D.~Acosta {\it et al.} (CDF Collaboration), {\it Phys. Rev.} {\bf D69}, 
052003 (2004); 
D.~Acosta {\it et al.} (CDF Collaboration),
{\it Phys. Rev.} {\bf D71}, 012005 (2005).

\bibitem{alldzero}
B.~Abbott~{\it et al.} (D\O~Collaboration), {\it Phys. Rev.} {\bf D63}, 031101 (2001);
V.~M.~Abazov~{\it et al.}~(D\O~Collaboration), {\it Phys. Lett.}
{\bf B517}, 282 (2001);  
V.~M.~Abazov~{\it et al.}~(D\O~Collaboration), {\it Phys. Lett.} {\bf B622}, 
265 (2005);
V.~M.~Abazov~{\it et al.} (D\O~Collaboration), hep-ex/0604020.

\bibitem{theo} B.~W.~Harris {\it et al.}, {\it Phys. Rev.} {\bf D66}, 054024 (2002); 
Z.~Sullivan,  {\it Phys. Rev.} {\bf D70}, 114012 (2004). 
J. Campbell, R.K. Ellis, F. Tramontano, {\it Phys. Rev.} {\bf D70}, 094012 (2004).

\bibitem{cdfnew} The CDF Collaboration, public conference note 8185, April 2006;
 M. B\"uhler, Search for Electroweak Single-Top Quark Production with 
 the CDF II Experiment, Diplomarbeit Universit\"at Karlsruhe, 
 FERMILAB-MASTERS-2006-02, August 2006.

\bibitem{d0new} The D\O \ Collaboration, public conference note 4871, July 2005.

\bibitem{secvtx} D. Acosta {\it et al.} (CDF Collaboration), {\it Phys. Rev.} 
{\bf D71}, 052003.

\bibitem{d0JLIP} S. Greder, Ph.D. thesis, Universit{\'e} Louis Pasteur, Strasbourg,
  IRES 05-006 ULP 4652 (2005).

\bibitem{d0Wprime} V.~M.~Abazov~{\it et al.}~(D\O~Collaboration),
{\it Phys. Lett.} {\bf B641}, 423 (2006).

\bibitem{comphep} E. Boos~{\it et al.}~({\sc comhep} collaboration), 
  {\it Nucl. Instrum. Methods} {\bf A534}, 463 (2004). 

\end{thebibliography}
\end{document}